\documentclass[12pt]{article}

\usepackage[cp1251]{inputenc}
\usepackage[russian]{babel}
\usepackage{graphicx,comment}
\usepackage{amsmath,amssymb,amsfonts,amsthm,comment}

\begin{document}

\noindent {\it Problems of Information Transmission},\\
\noindent vol. 53, no. 4, pp. 47--66, 2017.

\begin{center} {\bf M. V. Burnashev} \end{center}


\begin{center}
{\large\bf ON DETECTION OF GAUSSIAN STOCHACTIC SEQUENCES}
\footnote[1]{The research was carried out at the IITP RAS at the
expense of the Russian Foundation for Sciences
(project 14-50-00150).}
\end{center}

The problem of minimax detection of Gaussian random signal vector
in White Gaussian additive noise is considered. It is supposed that
an unknown vector $\boldsymbol{\sigma}$ of the signal vector
intensities belong to the given set ${\mathcal E}$. It is
investigated when it is possible to replace the set ${\mathcal E}$
by a smaller set ${\mathcal E}_{0}$ without loss of quality
(and, in particular, to replace it by a single point
$\boldsymbol{\sigma}_{0}$).

\vskip 0.7cm

\begin{center}
{\large\bf \S\;1. Inroduction}
\end{center}

{\bf 1. Simple hypotheses}. There are two simple hypotheses
$\mathcal H_{0}$ ("noise") and $\mathcal H_{1}$
("noise + stochastic signal") on observations
${\mathbf y} = (y_{1},\ldots,y_{n}) \in \mathbb{R}^{n}$:
\begin{equation}\label{intr0}
\begin{gathered}
\mathcal H_{0}: {\mathbf y} = \boldsymbol{\xi}, \\
\mathcal H_{1}: {\mathbf y} = {\mathbf s} + \boldsymbol{\xi},
\end{gathered}
\end{equation}
where $\boldsymbol{\xi} = (\xi_{1},\ldots,\xi_{n})$ -- independent
${\mathcal N}(0,1)$-Gaussian random variables, and
${\mathbf s} = (s_{1},\ldots,s_{n})$ -- independent on
$\boldsymbol{\xi}$, independent ${\mathcal N}(0,\sigma_{i}^{2})$,
$i=1,\ldots,n$-Gaussian random variables (i.e.
$\mathbf{E}(s_{i}^{2}) = \sigma_{i}^{2}$). Denote
$\boldsymbol{\sigma} = (\sigma_{1},\ldots,\sigma_{n})$, where
all $\sigma_{i} \geq 0$, and introduce functions
(those notations will also be used below)
\begin{equation}\label{defD}
\begin{gathered}
D(\boldsymbol{\sigma})= \sum_{i=1}^{n}\ln(1+\sigma_{i}^{2}), \quad
T(\boldsymbol{\sigma}) = \sum\limits_{i=1}^{n}
\dfrac{\sigma_{i}^{2}}{1+\sigma_{i}^{2}}, \quad
B(\boldsymbol{\sigma}) = 2\sum\limits_{j=1}^{n}
\frac{\sigma_{j}^{4}}{(1+\sigma_{j}^{2})^{2}}.
\end{gathered}
\end{equation}
Then for conditional probability densities we have
\begin{equation}\label{defp}
\begin{gathered}
p({\mathbf y}|\mathcal H_{0}) = (2\pi)^{-n/2}
e^{-\frac{1}{2}\sum\limits_{i=1}^{n}y_{i}^{2}}, \qquad
p({\mathbf y}|\boldsymbol{\sigma}) = (2\pi)^{-n/2}
e^{-\frac{1}{2}\sum\limits_{i=1}^{n}y_{i}^{2}/(1+\sigma_{i}^{2})-
\frac{1}{2}D(\boldsymbol{\sigma})}.
\end{gathered}
\end{equation}
Denote also
\begin{equation}\label{defr}
\begin{gathered}
r({\mathbf y},\boldsymbol{\sigma}) =
\ln \frac{p({\mathbf y}|\boldsymbol{\sigma})}
{p({\mathbf y}|\mathcal H_{0})} = \frac{1}{2}
\sum_{i=1}^{n}\frac{\sigma_{i}^{2}y_{i}^{2}}{1+\sigma_{i}^{2}} -
\frac{1}{2}D(\boldsymbol{\sigma}).
\end{gathered}
\end{equation}

The optimal solution of that problem of testing the simple
hypothesis $\mathcal H_{0}$ against the simple alternative
$\mathcal H_{1}$ (Neyman-Pierson criteria) \cite{Wald, Leman} has
the form
\begin{equation}\label{defcalA1}
{\mathbf y} \in \mathcal A(A,\boldsymbol{\sigma}) \Rightarrow
\mathcal H_{0}, \qquad
{\mathbf y} \not\in \mathcal A(A,\boldsymbol{\sigma}) \Rightarrow
\mathcal H_{1},
\end{equation}
where the set (ellipsoid) $\mathcal A(A,\boldsymbol{\sigma})$
\begin{equation}\label{defcalA}
\mathcal A(A,\boldsymbol{\sigma}) = \left\{{\mathbf y}:
\sum_{i=1}^{n}\frac{\sigma_{i}^{2}y_{i}^{2}}{1+\sigma_{i}^{2}}
\leq D(\boldsymbol{\sigma}) + A \right\}, \qquad
\boldsymbol{\sigma} = (\sigma_{1},\ldots,\sigma_{n}).
\end{equation}

The level $A$ of that test is determined by a given 1-st kind
error probability (``{\it false alarm probability}'')
$\alpha = \alpha(A,\boldsymbol{\sigma})$:
\begin{equation}\label{defalpha}
\alpha(A,\boldsymbol{\sigma}) =
\mathbf{P}({\mathbf y} \not\in \mathcal A|\mathcal H_{0})) =
\mathbf{P}\left(\sum_{i=1}^{n}\frac{\sigma_{i}^{2}\xi_{i}^{2}}
{1+\sigma_{i}^{2}} > D(\boldsymbol{\sigma}) +A\right).
\end{equation}

If hypothesis $\mathcal H_{1}$ is true then
$y_{i} = \xi_{i} + \sigma_{i}\eta_{i}
\sim \sqrt{1+\sigma_{i}^{2}}\,\eta_{i}$, where
$(\eta_{1},\ldots,\eta_{n})$ -- independent
${\mathcal N}(0,1)$-Gaussian random variables. Therefore
2-nd kind error probability (``{\it miss probability}'')
$\beta(A,\boldsymbol{\sigma})$ is defined by formula
\begin{equation}\label{defbeta}
\beta(A,\boldsymbol{\sigma}) =
\mathbf{P}({\mathbf y} \in \mathcal A|\mathcal H_{1}) =
\mathbf{P}\left(\sum_{i=1}^{n}\sigma_{i}^{2}\xi_{i}^{2}<
D(\boldsymbol{\sigma})+A\right).
\end{equation}

For a given value $\alpha$ denote by
$\beta(\alpha,\boldsymbol{\sigma})$ the minimum possible value
$\beta(A,\boldsymbol{\sigma})$ for optimal choice of level $A$
(according to formulas \eqref{defalpha}--\eqref{defbeta}).

Since $\mathbf{E}\xi_{i}^{2} = 1$, $i = 1,\ldots,n$, then due to
Large Numbers Law and \eqref{defalpha}--\eqref{defbeta} we get
that for sufficiently small $\alpha,\beta$ the value $A$ should
satisfy conditions
\begin{equation}\label{defA}
\sum_{i=1}^{n}\frac{\sigma_{i}^{2}}{1+\sigma_{i}^{2}} <
D(\boldsymbol{\sigma})+A < \sum_{i=1}^{n}\sigma_{i}^{2}.
\end{equation}
Below we assume satisfied both conditions \eqref{defA}. Note
that with decreasing the level $A$ the error probability
$\beta(A,\boldsymbol{\sigma})$ also decreases, but
the error probability $\alpha(A,\boldsymbol{\sigma})$ increases.
In particular, the case when the value $D(\boldsymbol{\sigma})+A$
is relatively close to the left side of the condition \eqref{defA}
will be interesting for us.

{\bf 2. Simple hypothesis against composite alternative}. Let
a set ${\mathcal E}$ of nonnegative vectors
$\boldsymbol{\sigma} = (\sigma_{1},\ldots,\sigma_{n})$ be given.
Assume that on the vector $\boldsymbol{\sigma}$, describing the
hypothesis $\mathcal H_{1}$ from \eqref{intr0} it is known only
that $\boldsymbol{\sigma} \in {\mathcal E}$, but the vector
$\boldsymbol{\sigma}$ itself is not known
(i.e. the hypothesis $\mathcal H_{1}$ is composite).

Similarly to \eqref{defcalA1}, for testing hypotheses
$\mathcal H_{0}$ and $\mathcal H_{1}$ we choose a decision region
${\mathcal A} \in \mathbb{R}^{n}$ such that
$$
{\mathbf y} \in \mathcal A \Rightarrow \mathcal H_{0}, \qquad
{\mathbf y} \not\in \mathcal A \Rightarrow \mathcal H_{1}.
$$
1-st kind and 2-nd kind error probabilities are defined by
formulas, respectively,
$$
\alpha({\mathcal A}) =
\mathbf{P}({\mathbf y} \not\in \mathcal A|\mathcal H_{0})
$$
and
$$
\beta({\mathcal A},{\mathcal E}) =
\mathbf{P}({\mathbf y} \in \mathcal A|\mathcal H_{1}) =
\sup\limits_{\boldsymbol{\sigma} \in {\mathcal E}}
\mathbf{P}({\mathbf y} \in \mathcal A|\boldsymbol{\sigma}).
$$
In other words, the minimax problem of testing hypotheses
$\mathcal H_{0}$ and $\mathcal H_{1}$ is considered.

Provided given 1-st kind error probability $\alpha$,
$0 < \alpha < 1$, we are interested in minimal possible 2-nd
kind error probability
\begin{equation}\label{defbeta3}
\beta(\alpha,{\mathcal E}) =
\inf\limits_{{\mathcal A}:\alpha({\mathcal A}) \leq \alpha}
\beta({\mathcal A},{\mathcal E})
\end{equation}
and corresponding decision region ${\mathcal A}(\alpha)$.

Without loss of generality we assume the set ${\mathcal E}$
closed and Lebeques measurable on $\mathbb{R}^{n}$. Formally
speaking, the optimal solution of the problem \eqref{defbeta3}
of minimax testing of hypotheses $\mathcal H_{0}$ and
$\mathcal H_{1}$ is described in Wald's general theory of
statistical decisions \cite{Wald}. For that solution we need to
find the ``{\it least favorable}'' prior distribution
$\pi_{\rm lf}(d{\mathcal E})$ on ${\mathcal E}$, replace the
composite hypothesis $\mathcal H_{1}$ by simple hypothesis
$\mathcal H_{1}(\pi_{\rm lf})$, and then to investigate
characteristics of corresponding Neyman-Pierson criteria for
testing simple hypotheses $\mathcal H_{0}$ and
$\mathcal H_{1}(\pi_{\rm lf})$. Unfortunately, all that can be
done only in some very special cases. Therefore it is natural
to separate cases, when that ``least favorable'' prior distribution
on ${\mathcal E}$ has the simplest form (for example, it is
concentrated in one point from ${\mathcal E}$).

Clearly, for the value $\beta(\alpha,{\mathcal E})$ the lower
bound holds
\begin{equation}\label{defbeta4}
\beta(\alpha,{\mathcal E}) \geq
\sup\limits_{\boldsymbol{\sigma} \in {\mathcal E}}
\beta(\alpha,\boldsymbol{\sigma}).
\end{equation}
The function $\beta(\alpha,\boldsymbol{\sigma})$,
$\alpha \in [0,1]$, $\boldsymbol{\sigma} \in \mathbb{R}^{n}_{+}$
is continuous on both arguments. Since the set
${\mathcal E} \in \mathbb{R}^{n}_{+}$ supposed to be closed then
there exists $\boldsymbol{\sigma}_{0} =
\boldsymbol{\sigma}_{0}({\mathcal E},\alpha) \in {\mathcal E}$,
such that
$$
\beta(\alpha,\boldsymbol{\sigma}_{0}) =
\sup\limits_{\boldsymbol{\sigma} \in {\mathcal E}}
\beta(\alpha,\boldsymbol{\sigma}).
$$
First, we are interested for what kind of ${\mathcal E}$ the
``least favorable'' prior distribution is
concentrated in the point $\boldsymbol{\sigma}_{0}$ and then
the following equality holds
\begin{equation}\label{defbeta6}
\beta(\alpha,{\mathcal E}) =
\beta(\alpha,\boldsymbol{\sigma}_{0}).
\end{equation}

If for the set ${\mathcal E}$ the equality \eqref{defbeta6} holds
then without any loss of detection quality we may replace
the composite hypothesis $\mathcal H_{1} = \{{\mathcal E}\}$ by
the simple hypothesis $\mathcal H_{1} = \boldsymbol{\sigma}_{0}$
and the optimal solution \eqref{defcalA1}--\eqref{defcalA} for
the simple hypothesis $\mathcal H_{1} = \boldsymbol{\sigma}_{0}$
remains optimal (in minimax sense) for the composite hypothesis
$\mathcal H_{1} = \{{\mathcal E}\}$ as well (see similar question
for shifts of measures \cite{Bur79}). Some sufficient conditions
for having the equality \eqref{defbeta6} are given below in
$\S\,3$. Of course, those conditions set rather strong
limitations on the set ${\mathcal E}$.

Earlier, it is shown in $\S\,2$ that sometimes it is possible
without any loss of detection quality to replace the set
${\mathcal E}$ by a smaller set ${\mathcal E}_{0}$ (i.e. to make
a reduction of the set ${\mathcal E}$).

Usually in the problem considered the probability
$\beta(\alpha,{\mathcal E})$ should be very small. For that
reason often instead of the strong condition \eqref{defbeta6}
its simpler asymptotic analogue is investigated, comparing
exponents of error probabilities (see, for example,
\cite{ZhangPoor11}). In that case we are interested in validity
of a weaker condition:
\begin{equation}\label{compar1a}
\ln \beta(\alpha,{\mathcal E}) =
\ln \beta(\alpha,\boldsymbol{\sigma}) +
o\left(\ln\beta(\alpha,\boldsymbol{\sigma})\right),
\qquad |\ln \beta(\alpha,\boldsymbol{\sigma})| \to \infty.
\end{equation}
It will be shown below that the condition \eqref{compar1a}
holds under a weaker restrictions on the set ${\mathcal E}$,
than in the case of the condition \eqref{defbeta6}.

Note that if for the set ${\mathcal E}$ asymptotic equality
\eqref{compar1a} holds, it does not mean that optimal
solution \eqref{defcalA1}--\eqref{defcalA} for simple hypothesis
$\mathcal H_{1} = \boldsymbol{\sigma}_{0}$ remains optimal for
composite hypothesis $\mathcal H_{1} = \{{\mathcal E}\}$
as well. Probably, it will be necessary to use another test.
Some sufficient conditions for having equality \eqref{compar1a}
and corresponding test are described in $\S\,4$.

Since in the problem considered the probability
$\beta(\alpha,{\mathcal E})$ usually should be very small,
in the paper large deviations for the value
$\beta(\alpha,{\mathcal E})$ (i.e. its logarithmic asymptotics
as $n \to \infty$) is also investigated. In $\S\,4$ for that
asymptotics upper bounds and in Appendix lower bounds are
obtained (from which the exact logarithmic asymptotics of
$\beta(\alpha,{\mathcal E})$ as $n \to \infty$ follows).
In $\S\,5$ similar upper bounds for the value
$\alpha(A,\boldsymbol{\sigma})$ are derived. If the value
$\alpha(A,\boldsymbol{\sigma})$ is not too small, then in order
to have completeness in $\S\,5.1$ it is investigated using
the Central Limit Theorem and Berry-Esseen inequality, which give
a more accurate estimates.

In $\S\,6$ a special example is considered. Some useful estimates
for large deviations of the distribution $\chi^{2}$, used in
the paper, are given in Appendix.

All formulas in the paper are, essentially, non-asymptotic.
All remaining terms can always be estimated.

Below, as usual, $\boldsymbol{\sigma} \leq \boldsymbol{\lambda}$
means $\sigma_{i} \leq \lambda_{i}$, $i=1,\ldots,n$.

\begin{center}
{\large\bf \S\,2. Reduction of the set ${\mathcal E}$}
\end{center}

We show that sometimes without any loss of detection quality
it is possible to replace the set ${\mathcal E}$ by a smaller
set ${\mathcal E}_{0}$. Define such set
${\mathcal E}_{0} = {\mathcal E}_{0}({\mathcal E})$ as any set
having the following property:
\begin{equation}\label{defe0}
\begin{gathered}
\mbox{for any $\boldsymbol{\sigma} \in {\mathcal E}$ there exusts
$\boldsymbol{\sigma}_{0} \in {\mathcal E}_{0}$ with
$\boldsymbol{\sigma}_{0} \leq \boldsymbol{\sigma}$}.
\end{gathered}
\end{equation}
If the set ${\mathcal E}$ is closed (it is assumed in the paper),
then ${\mathcal E}_{0} \subseteq {\mathcal E}$. Generally, the set
${\mathcal E}_{0}$ can be chosen non-uniquely.

It is shown below that for any Bayes criteria of testing a
simple hypothesis $\mathcal H_{0}$ against a composite
alternative $\mathcal H_{1} = \{{\mathcal E}\}$ the set
${\mathcal E}$ can be replaced by the set ${\mathcal E}_{0}$
without any loss of quality. It remains valid for
likelihood ratio criteria as well. In one-dimensional case
those properties are similar to the case of distributions with
monotone likelihood ratio \cite[Ch. 3.9]{Leman}.

The aim of the set ${\mathcal E}_{0}$ introduction is to decrease
(if possible) the set ${\mathcal E}$ and so to simplify the test used.

{\bf 1. Bayes criteria}. Consider a Bayes criteria with a prior
distribution
$\pi(d{\mathcal E})$ on ${\mathcal E}$ and corresponding decision
set ${\mathcal A} \in \mathbb{R}^{n}$
(${\mathbf y} \in \mathcal A \Rightarrow \mathcal H_{0}$,
${\mathbf y} \not\in \mathcal A \Rightarrow \mathcal H_{1}$)
of the form
\begin{equation}\label{defAA}
\begin{gathered}
{\mathcal A}= {\mathcal A}(A)= \left\{{\mathbf y}:
r({\mathbf y},{\mathcal E},\pi) \leq A\right\},
\end{gathered}
\end{equation}
where (see \eqref{defp}, \eqref{defr})
$$
\begin{gathered}
p({\mathbf y}|\mathcal H_{1},\pi) =
\int\limits_{\boldsymbol{\sigma} \in {\mathcal E}}
p({\mathbf y}|\boldsymbol{\sigma})\pi(d{\mathcal E})
= (2\pi)^{-n/2}\int\limits_{\boldsymbol{\sigma} \in {\mathcal E}}
e^{-\frac{1}{2}
\sum\limits_{i=1}^{n}y_{i}^{2}/(1+\sigma_{i}^{2})-
\frac{1}{2}D(\boldsymbol{\sigma})}\pi(d{\mathcal E})
\end{gathered}
$$
and
$$
\begin{gathered}
r({\mathbf y},{\mathcal E},\pi) =
\ln\frac{p({\mathbf y}|\mathcal H_{1},\pi)}
{p({\mathbf y}|\mathcal H_{0})} =
\ln\int\limits_{\boldsymbol{\sigma} \in {\mathcal E}}
e^{\frac{1}{2}\sum\limits_{i=1}^{n}
y_{i}^{2}\sigma_{i}^{2}/(1+\sigma_{i}^{2})-
\frac{1}{2}D(\boldsymbol{\sigma})}
\pi(d{\mathcal E}).
\end{gathered}
$$
Then ${\mathcal A}$ -- convex set in $\mathbb{R}^{n}$, and if
${\mathbf y} = (y_{1},\ldots,y_{n})\in {\mathcal A}$, then all
$(\pm y_{1},\ldots,\pm y_{n})$ belong to
${\mathcal A}$, i.e. the set ${\mathcal A}$ is symmetric with
respect to any coordinate axis or plane. In particular, such
${\mathcal A}$ is also centrally symmetric set
(i.e. if ${\mathbf y} \in {\mathcal A}$, then
$(-{\mathbf y}) \in {\mathcal A}$).

Assume that for ${\mathbf y}= {\mathbf s} + \boldsymbol{\xi}$,
$\boldsymbol{\sigma} \in {\mathcal E}$ from \eqref{intr0}
a Bayes criteria with a prior distribution
$\pi(d{\mathcal E})$ на ${\mathcal E}_{0}$ is used and
${\mathcal A} \in \mathbb{R}^{n}$ of the form \eqref{defAA}
is the corresponding decision region. Assume also that for
2-nd kind error probability and some $\beta \geq 0$ we have
\begin{equation}\label{defbetabei}
\beta({\mathcal A},\boldsymbol{\sigma}_{0}) = \mathbf{P}
({\mathbf y} \in \mathcal A|\boldsymbol{\sigma}_{0}) =
\mathbf{P}\left\{r({\mathbf y},{\mathcal E}_{0},\pi) \leq A
|\boldsymbol{\sigma}_{0}\right\} \leq \beta, \qquad
\boldsymbol{\sigma}_{0} \in {\mathcal E}_{0}.
\end{equation}
Show that the inequality \eqref{defbetabei} remains valid for
any $\boldsymbol{\sigma} \in {\mathcal E}$, i.e.
\begin{equation}\label{defbetabei1}
\beta({\mathcal A},\boldsymbol{\sigma}) = \mathbf{P}
({\mathbf y} \in \mathcal A|\boldsymbol{\sigma}) =
\mathbf{P}\left\{r({\mathbf y},{\mathcal E}_{0},\pi) \leq A
|\boldsymbol{\sigma}\right\} \leq \beta, \qquad
\boldsymbol{\sigma} \in {\mathcal E}.
\end{equation}

In other words, for any Bayes criteria extension of the set
${\mathcal E}_{0}$ up to the set ${\mathcal E}$ does not
increase 2-nd kind error probability (1-st kind
error probability $\alpha({\mathcal A})$ does not change).
In particular, since ${\mathcal E}_{0} \subseteq {\mathcal E}$,
we get
\begin{equation}\label{defbetabei2}
\beta(\alpha, {\mathcal E}_{0}) = \beta(\alpha, {\mathcal E}),
\qquad 0 \leq \alpha \leq 1.
\end{equation}

We prove the relation \eqref{defbetabei1}. Let
$\boldsymbol{\sigma} \in {\mathcal E}$, but
$\boldsymbol{\sigma} \not\in {\mathcal E}_{0}$. Then there exists
$\boldsymbol{\sigma}_{0} \in {\mathcal E}_{0}$ with
$\boldsymbol{\sigma}_{0} < \boldsymbol{\sigma}$. Let
${\mathbf s}_{0}$ - Gaussian ``signal'' in \eqref{intr0} in the
case of $\boldsymbol{\sigma}_{0}$. Then in the case of
$\boldsymbol{\sigma}$ such ``signal'' ${\mathbf s}$ has the form
${\mathbf s} = {\mathbf s}_{0} + \boldsymbol{\eta}$, where
$\boldsymbol{\eta}$ -- independent of ${\mathbf s}_{0}$
Gaussian random vector. The inequality \eqref{defbetabei1}
follows from the following auxiliary result
(the set ${\mathcal A}$ satisfies its conditions).

L e m m a\, 1. {\it Let ${\mathcal B}\in \mathbb{R}^{n}$ -
a convex set, such that if
${\mathbf y} = (y_{1},\ldots,y_{n})\in {\mathcal B}$, then all
points of the form $(\pm y_{1},\ldots,\pm y_{n})$ belong to
${\mathcal B}$. Let also $\boldsymbol{\xi}, \boldsymbol{\eta}$ --
independent zero mean Gaussian vectors, consisting of
independent (probably, with different distributions) components.
Then }
\begin{equation}\label{ineq0}
\begin{gathered}
\mathbf{P}(\boldsymbol{\xi} + \boldsymbol{\eta} \in \mathcal B)
\leq \mathbf{P}(\boldsymbol{\xi} \in \mathcal B).
\end{gathered}
\end{equation}

P r o o f. If $n=1$, then ${\mathcal B} = [-a,a]$,
$a >0$ and, clearly, the inequality \eqref{ineq0} holds.
Let $n = 2$ and vectors $(\xi_{1},\xi_{2})$ and
$(\xi_{1}+\eta_{1},\xi_{2}+\eta_{2})$ are compared. Compare first
vectors $(\xi_{1},\xi_{2})$ and $(\xi_{1}+\eta_{1},\xi_{2})$.
Denote
$$
\begin{gathered}
{\mathcal B}_{x} = \{{\mathbf y}\in {\mathcal B}:y_{2}=x\}
\in \mathbb{R}^{1}.
\end{gathered}
$$
Due to assumptions of Lemma, for any $x$ we have
${\mathcal B}_{x} = [-a(x),a(x)]$, $a(x) > 0$. Therefore fox fixed
$\xi_{2}$ the problem reduces to the case $n=1$ and
\begin{equation}\label{ineq01}
\begin{gathered}
\mathbf{P}\{\xi_{1}+\eta_{1} \in \mathcal B_{\xi_{2}}\}
\leq \mathbf{P}\{\xi_{1} \in \mathcal B_{\xi_{2}}\}
\end{gathered}
\end{equation}
and then
\begin{equation}\label{ineq02}
\begin{gathered}
\mathbf{P}\{(\xi_{1}+\eta_{1},\xi_{2}) \in \mathcal B\}
\leq \mathbf{P}\{(\xi_{1},\xi_{2})\in \mathcal B\}.
\end{gathered}
\end{equation}
Compare now vectors $(\xi_{1}+\eta_{1},\xi_{2})$ and
$(\xi_{1}+\eta_{1},\xi_{2}+\eta_{2})$. Similarly to
\eqref{ineq01} and \eqref{ineq02} we get
$$
\begin{gathered}
\mathbf{P}\{\xi_{2}+\eta_{2} \in \mathcal B_{\xi_{1}+\eta_{1}}\}
\leq \mathbf{P}\{\xi_{2} \in \mathcal B_{\xi_{1}+\eta_{1}}\}
\end{gathered}
$$
and
\begin{equation}\label{ineq04}
\begin{gathered}
\mathbf{P}\{(\xi_{1}+\eta_{1},\xi_{2}+\eta_{2}) \in \mathcal B\}
\leq \mathbf{P}\{(\xi_{1}+\eta_{1},\xi_{2})\in \mathcal B\}.
\end{gathered}
\end{equation}
Then by \eqref{ineq02} and \eqref{ineq04} the inequality
\eqref{ineq0} follows for $n=2$. Similarly, the case $n=3$
reduces to the case $n=2$ and so on. It proves the inequality
\eqref{ineq0} for any $n$. $\qquad \triangle$

{\bf 2. Likelihood ratio criteria}. For any function
$A(\boldsymbol{\sigma})$ the critical region
${\mathcal A}_{ML}(A,{\mathcal E})$ of that criteria is defined
by the relation
\begin{equation}\label{defmaxrat1}
{\mathcal A}_{ML}(A,{\mathcal E}) =\left\{{\mathbf y}:
\sup_{\boldsymbol{\sigma} \in {\mathcal E}}\left[
2r({\mathbf y},\boldsymbol{\sigma})-
A(\boldsymbol{\sigma})\right] \leq 0 \right\}
\end{equation}
and then ${\mathbf y} \in {\mathcal A}_{ML}(A,{\mathcal E})
\Rightarrow \mathcal H_{0}$,
${\mathbf y} \not\in {\mathcal A}_{ML}(A,{\mathcal E})
\Rightarrow \mathcal H_{1}$.

Show that without any loss of quality we may replace  the set
${\mathcal E}$ in \eqref{defmaxrat1} by smaller set
${\mathcal E}_{0}$ (see \eqref{defe0}), i.e. to use the criteria:
\begin{equation}\label{defmaxrat1a}
{\mathcal A}_{MLR}(A,{\mathcal E}) =\left\{{\mathbf y}:
\sup_{\boldsymbol{\sigma} \in {\mathcal E}_{0}}\left[
2r({\mathbf y},\boldsymbol{\sigma})-
A(\boldsymbol{\sigma})\right] \leq 0 \right\},
\end{equation}
keeping the same decision making method. In other words, for
likelihood ratio criteria expansion of the set ${\mathcal E}_{0}$
up to the set ${\mathcal E}$ does not increase the 2-nd kind
error probability (the 1-st kind error probability
$\alpha({\mathcal A})$ does not change).

Indeed, if $\boldsymbol{\sigma} \in {\mathcal E}$, but
$\boldsymbol{\sigma} \not\in {\mathcal E}_{0}$, then there exists
$\boldsymbol{\sigma}_{0} \in {\mathcal E}_{0}$ with
$\boldsymbol{\sigma}_{0} < \boldsymbol{\sigma}$. Using the
definition \eqref{defmaxrat1a} and formulas
\eqref{deflambda} and \eqref{defa1} below, we have
\begin{equation}\label{defbeta4a}
\begin{gathered}
\beta(A,\boldsymbol{\sigma}) =
\mathbf{P}\left\{\sup_{\boldsymbol{\lambda} \in {\mathcal E}_{0}}
\left[2r({\mathbf y},\boldsymbol{\lambda})-A(\boldsymbol{\lambda})
\right]\leq 0|\boldsymbol{\sigma}\right\} =  \\
= \mathbf{P}\left\{\sup_{\boldsymbol{\lambda} \in {\mathcal E}_{0}}
\left[\sum_{i=1}^{n}\frac{\lambda_{i}^{2}(1+\sigma_{i}^{2})
\eta_{i}^{2}}{1+\lambda_{i}^{2}} -D(\boldsymbol{\lambda})-
A(\boldsymbol{\lambda})\right]\leq 0\right\} \leq \\
\leq \mathbf{P}\left\{\sup_{\boldsymbol{\lambda} \in {\mathcal E}_{0}}
\left[\sum_{i=1}^{n}\frac{\lambda_{i}^{2}(1+\sigma_{0i}^{2})
\eta_{i}^{2}}{1+\lambda_{i}^{2}} -D(\boldsymbol{\lambda}) -
A(\boldsymbol{\lambda})\right]\leq 0\right\} = \\
=\mathbf{P}\left\{\sup_{\boldsymbol{\lambda} \in {\mathcal E}_{0}}
\left[2r({\mathbf y},\boldsymbol{\lambda})-A(\boldsymbol{\lambda})
\right]\leq 0|\boldsymbol{\sigma}_{0}\right\} =
\beta(A,\boldsymbol{\sigma}_{0}).
\end{gathered}
\end{equation}

Results \eqref{defbetabei1} and \eqref{defbeta4a} obtained can be
formulated as follows.

P r o p o s i t i o n \,1. {\it Consider the minimax problem of
testing a simple hypothesis $\mathcal H_{0}$ against a
composite alternative $\mathcal H_{1} = \{{\mathcal E}_{0}\}$ and
let ${\mathcal E}_{0} \subseteq {\mathcal E}$. If for the set
${\mathcal E}$ the condition \eqref{defe0} is satisfied then
for any Bayes criteria and the likelihood ratio criteria
the 1-st kind and the 2-nd kind error probabilities
do not change if the set ${\mathcal E}_{0}$ is replaced by the
set ${\mathcal E}$. In particular, the equality
\eqref{defbetabei2} holds}.

{\it Remark} 1. It seems that it would be more natural in
Proposition 1 to start with a set  ${\mathcal E}$ and to replace it
by a set ${\mathcal E}_{0} \subseteq {\mathcal E}$. But in that
case it would be necessary to describe ``projections'' of
Bayes criteria from ${\mathcal E}$ on ${\mathcal E}_{0}$.

{\it Remark} 2. Similar to $\mathcal{E}_0$ ``reduced'' sets
$\mathop{\rm red}_1S$
and $\mathop{\rm red}_2S$ have been introduced earlier in \cite{Bur79},
where Gaussian measures differed from each other only by shifts.
From analytical viewpoint, various convexity properties with respect
to shifts of Gaussian measures were very useful in \cite{Bur79}.
For example, due to them the set $\mathop{\rm red}_1S$ had very simple
and natural form. Unfortunately, the author does not know similar
convexity properties concerning variances of Gaussian measures and
for that reason only certain monotonicity properties have been
used (what is less productive).

\begin{center}
{\large\bf \S\,3. Exact equality \eqref{defbeta6}}
\end{center}

{\bf 1}. The formula \eqref{defbeta6} has also another equivalent
interpretation. Assume that initially we know that in the
hypothesis $\mathcal H_{1}$ the ``signal'' is a certain
$\boldsymbol{\sigma}$ and therefore we use the optimal solution
\eqref{defcalA1}--\eqref{defcalA} for that $\boldsymbol{\sigma}$.
Assume additionally that in fact the ``signal'' in the hypothesis
$\mathcal H_{1}$ may also take another values
$\boldsymbol{\lambda}$ from a set ${\mathcal E}$. For what
${\mathcal E}$ the solution \eqref{defcalA1}--\eqref{defcalA}
(oriented only on $\boldsymbol{\sigma}$) remains optimal for the
set ${\mathcal E}$ as well ?

If $\boldsymbol{\sigma}$ is replaced by $\boldsymbol{\lambda}$
and decision \eqref{defcalA1}--\eqref{defcalA} is used, then
the 1-st kind error probability $\alpha$ does not change.
Therefore it is necessary to check only how the 2-nd kind error
probability $\beta_{\boldsymbol{\sigma}}(A,\boldsymbol{\lambda})$
may change
\begin{equation}\label{deflambda}
\begin{gathered}
\beta_{\boldsymbol{\sigma}}(A,\boldsymbol{\lambda}) =
\mathbf{P}({\mathbf y} \in \mathcal A|\boldsymbol{\lambda}) =
\mathbf{P}\left(\sum_{i=1}^{n}
\frac{\sigma_{i}^{2}(\xi_{i} + s_{i})^{2}}{1+\sigma_{i}^{2}}-
D(\boldsymbol{\sigma}) < A|\boldsymbol{\lambda}\right) = \\
= \mathbf{P}\left(\sum_{i=1}^{n}\nu_{i}^{2}
\xi_{i}^{2}-D(\boldsymbol{\sigma})< A\right),
\end{gathered}
\end{equation}
since $(\xi_{i} + s_{i})^{2} = (1+\lambda_{i}^{2})\eta_{i}^{2}$,
$i=1,\ldots,n$ and where
\begin{equation}\label{defa1}
\nu_{i}^{2} = \frac{\sigma_{i}^{2}(1+\lambda_{i}^{2})}
{1+\sigma_{i}^{2}} = \sigma_{i}^{2} +
\frac{\sigma_{i}^{2}(\lambda_{i}^{2} - \sigma_{i}^{2})}
{1+\sigma_{i}^{2}}, \qquad i=1,\ldots,n,
\end{equation}
and $\{\eta_{i}\}$ - independent ${\mathcal N}(0,1)$-Gaussian
random variables.

If for any $\boldsymbol{\lambda} \in {\mathcal E}$ and $A$ the
following inequality holds
($\boldsymbol{\nu} = (\nu_{1},\ldots,\nu_{n})$ is defined in
\eqref{defa1})
\begin{equation}\label{compar1}
\begin{gathered}
\beta_{\boldsymbol{\sigma}}(A,\boldsymbol{\lambda}) =
\mathbf{P}\left(\sum_{i=1}^{n}\nu_{i}^{2}\xi_{i}^{2}-
D(\boldsymbol{\sigma}) < A\right) \leq
\mathbf{P}\left(\sum_{i=1}^{n}\sigma_{i}^{2}\xi_{i}^{2} -
D(\boldsymbol{\sigma}) < A\right)=
\beta(A,\boldsymbol{\sigma}),
\end{gathered}
\end{equation}
then
$$
\beta(A,{\mathcal E})  \leq
\sup_{\boldsymbol{\lambda} \in {\mathcal E}}
\beta_{\boldsymbol{\sigma}}(A,\boldsymbol{\lambda}) \leq
\beta(A,\boldsymbol{\sigma})
$$
and therefore the formula \eqref{defbeta6} is valid.

Some results showing validity of the inequality \eqref{compar1}
for certain $\boldsymbol{\sigma}, \boldsymbol{\nu},A$
can be found, for example, in \cite{Ponom85, Bak95, Bur16}.

In order to have
$\beta_{\boldsymbol{\sigma}}(A,\boldsymbol{\lambda}) \leq
\beta(A,\boldsymbol{\sigma})$ for any $A$ (see \eqref{compar1}),
it is necessary, at least, to have (comparing of expectations)
$$
\sum_{i=1}^{n}\nu_{i}^{2}- \sum_{i=1}^{n}\sigma_{i}^{2} =
\sum_{i=1}^{n}\frac{\sigma_{i}^{2}(\lambda_{i}^{2}-\sigma_{i}^{2})}
{1+\sigma_{i}^{2}} \geq 0.
$$

Comparing \eqref{defbeta}, \eqref{deflambda} and \eqref{defa1},
we get simple

P r o p o s i t i o n \,2. 1) {\it If
$\boldsymbol{\sigma} \leq \boldsymbol{\lambda}$, then
$\beta(A,\boldsymbol{\lambda}) \leq \beta(A,\boldsymbol{\sigma})$
and $\beta_{\boldsymbol{\sigma}}(A,\boldsymbol{\lambda}) \leq
\beta(A,\boldsymbol{\sigma})$ for any $A$.}

2) {\it If
$\boldsymbol{\sigma} \leq \boldsymbol{\lambda}$ for any
$\boldsymbol{\lambda} \in {\mathcal E}$, then
$\beta(\alpha,{\mathcal E}) =\beta(\alpha,\boldsymbol{\sigma})$
for any $\alpha$. }

{\bf 2}. As an example consider the following result, which is the
part of lemma 1 from \cite{Bur16}.

L e m m a \ 2. {\it Assume that the set of indices
$I = \{1,2,\ldots,n\}$ of vectors
$\boldsymbol{\sigma},\boldsymbol{\lambda}$ can be partitioned
in $k \geq 1$ groups $I_{1},\ldots,I_{k}$, such that
$I = \bigcup\limits_{j=1}^{k}I_{j}$, $I_{i}\cap I_{j}= \emptyset$,
$i \neq j$, and the following conditions are fulfilled
$$
\begin{gathered}
\sigma_{i} \leq \lambda_{0,j}, \qquad i \in I_{j}, \quad
j=1,\ldots,k,
\end{gathered}
$$
where
$$
\begin{gathered}
\lambda_{0,j}=
\left(\prod\limits_{i \in I_{j}}\lambda_{i}\right)^{1/|I_{j}|}.
\end{gathered}
$$
Then $\beta(A,\boldsymbol{\lambda}) \leq
\beta(A,\boldsymbol{\sigma})$ for any $A$.}

E x a m p l e \, 1. Let for given $D > 0$
$$
{\mathcal E} = \left\{\boldsymbol{\lambda} \geq {\mathbf 0}:
\prod_{i=1}^{n}\left(1+\lambda_{i}^{2}\right) \geq
\left(1+D^{2}\right)^{n}\right\}.
$$
Then from the formula \eqref{defa1} and Lemma 2 with $k=1$
it follows that the set ${\mathcal E}$ can be replaced
(without loss of quality) by single point
$\boldsymbol{\sigma}_{0} = (D,\ldots,D) \in {\mathcal E}$
(in the sense of exact equality \eqref{defbeta6}).

\begin{center}
{\large\bf \S\,4. Asymptotic equality \eqref{compar1a}. Large
deviations for $\beta(A,\boldsymbol{\sigma})$ and
$\beta_{\boldsymbol{\sigma}}(A,\boldsymbol{\lambda})$}
\end{center}

Consider conditions when the equality \eqref{compar1a} holds.
For that purpose we investigate the logarithmic asymptotics
of probabilities $\beta(A,\boldsymbol{\sigma})$ and
$\beta_{\boldsymbol{\sigma}}(A,\boldsymbol{\lambda})$
as $n \to \infty$.

{\bf 1. Large deviations. Upper bounds}. Since for
$\xi \sim {\cal N}(0,1)$
\begin{equation}\label{ineq18}
\mathbf{E}e^{a(\xi + b)^{2}} = \frac{1}{\sqrt{1-2a}}
\exp\left\{\frac{2ab^{2}}{1-2a}\right\}, \qquad a < 1/2, \qquad
b \in \mathbb{R}^{1},
\end{equation}
then using exponential Chebychev inequality for any $u \geq 0$
we have
\begin{equation}\label{ineq1a}
\begin{gathered}
\beta(A,\boldsymbol{\sigma}) = \mathbf{P}
\left(\sum_{i=1}^{n}\sigma_{i}^{2}\xi_{i}^{2} <
D(\boldsymbol{\sigma})+A\right) \leq
e^{u[D(\boldsymbol{\sigma})+ A]/2}
\mathbf{E}e^{-u \sum\limits_{i=1}^{n}\sigma_{i}^{2}\xi_{i}^{2}/2} =
e^{-g_{\boldsymbol{\sigma}}(u)},
\end{gathered}
\end{equation}
where
\begin{equation}\label{ineq1b}
\begin{gathered}
2g_{\boldsymbol{\sigma}}(u) = \sum\limits_{i=1}^{n}
\ln(1+u \sigma_{i}^{2}) -u[D(\boldsymbol{\sigma})+ A], \quad
2g_{\boldsymbol{\sigma}}'(u) = \sum\limits_{i=1}^{n}
\frac{\sigma_{i}^{2}}{1+u\sigma_{i}^{2}}-D(\boldsymbol{\sigma})-A,
\\ \qquad
g_{\boldsymbol{\sigma}}''(u) < 0.
\end{gathered}
\end{equation}
Since both conditions \eqref{defA} supposed to be fulfilled, then
$g_{\boldsymbol{\sigma}}'(0) > 0$ and
$g_{\boldsymbol{\sigma}}'(1) < 0$. Therefore
$\max\limits_{u \geq 0}g_{\boldsymbol{\sigma}}(u)$ is attained for
$0 < u_{0} < 1$, which is determined by the equation
$g_{\boldsymbol{\sigma}}'(u_{0}) = 0$, i.e.
\begin{equation}\label{approx3a1}
\sum\limits_{i=1}^{n}
\frac{\sigma_{i}^{2}}{1+u_{0}\sigma_{i}^{2}} =
D(\boldsymbol{\sigma})+A.
\end{equation}
Then from \eqref{ineq1a} and \eqref{ineq1b} we get
\begin{equation}\label{approx3a3}
\beta(A,\boldsymbol{\sigma}) \leq
e^{-g_{\boldsymbol{\sigma}}(u_{0})},
\end{equation}
where
\begin{equation}\label{approx3a41}
g_{\boldsymbol{\sigma}}(u_{0})= \max\limits_{u \geq 0}
g_{\boldsymbol{\sigma}}(u).
\end{equation}
Provided certain conditions (see Appendix, point 3) it is exact
logarithmic asymptotics of the value $\beta(A,\boldsymbol{\sigma})$
as $n \to \infty$.

Similarly, from \eqref{deflambda} and \eqref{defa1} for any
$v \geq 0$ we have
$$
\begin{gathered}
\beta_{\boldsymbol{\sigma}}(A,\boldsymbol{\lambda}) = \mathbf{P}
\left(\sum\limits_{i=1}^{n}\nu_{i}^{2}\xi_{i}^{2} <
D(\boldsymbol{\sigma})+A\right)
\leq e^{v[D(\boldsymbol{\sigma})+ A]/2}\prod\limits_{i=1}^{n}
\frac{1}{\sqrt{1+v \nu_{i}^{2}}} =
e^{-g_{\boldsymbol{\sigma}}(v,\boldsymbol{\lambda})},
\end{gathered}
$$
where values $\{\nu_{i}^{2}\}$ are defined in \eqref{defa1} and
$$
\begin{gathered}
2g_{\boldsymbol{\sigma}}(v,\boldsymbol{\lambda}) =
\sum\limits_{i=1}^{n}\ln(1+v \nu_{i}^{2}) -
v[D(\boldsymbol{\sigma})+A], \\
2g'_{\boldsymbol{\sigma}}(v,\boldsymbol{\lambda}) =
\sum\limits_{i=1}^{n}
\frac{\nu_{i}^{2}}{1+v\nu_{i}^{2}}-D(\boldsymbol{\sigma})-A,
\qquad g''_{\boldsymbol{\sigma}}(v,\boldsymbol{\lambda}) < 0.
\end{gathered}
$$
Then
\begin{equation}\label{approx3cd}
\beta_{\boldsymbol{\sigma}}(A,\boldsymbol{\lambda}) \leq
e^{-g_{\boldsymbol{\sigma}}(v_{0},\boldsymbol{\lambda})},
\end{equation}
where
\begin{equation}\label{approx3cd1}
g_{\boldsymbol{\sigma}}(v_{0},\boldsymbol{\lambda}) =
\max\limits_{v \geq 0}
g_{\boldsymbol{\sigma}}(v,\boldsymbol{\lambda}).
\end{equation}
There is a sense to consider only $\boldsymbol{\lambda}$ such that
$g'_{\boldsymbol{\sigma}}(0,\boldsymbol{\lambda})
= \sum\limits_{i=1}^{n}\nu_{i}^{2} - D(\boldsymbol{\sigma})-A > 0$
(otherwise $v_{0} = 0$). Then $\max\limits_{v \geq 0}
g_{\boldsymbol{\sigma}}(v,\boldsymbol{\lambda})$ is attained for
$v_{0} > 0$, which is determined by the equation
$$
\sum\limits_{i=1}^{n}\frac{\nu_{i}^{2}}{1+v_{0}\nu_{i}^{2}} =
\sum\limits_{i=1}^{n}\frac{\sigma_{i}^{2}(1+\lambda_{i}^{2})}
{1+\sigma_{i}^{2}+v_{0}\sigma_{i}^{2}(1+\lambda_{i}^{2})}=
D(\boldsymbol{\sigma})+A.
$$

If estimates \eqref{approx3a3}--\eqref{approx3a41} and
\eqref{approx3cd}--\eqref{approx3cd1} give the right logarithmic
asymptotics (as $n \to \infty$) of values
$\beta(A,\boldsymbol{\sigma})$ and
$\beta_{\boldsymbol{\sigma}}(A,\boldsymbol{\lambda})$, then
$g_{\boldsymbol{\sigma}}(u_{0}) -
g_{\boldsymbol{\sigma}}(v_{0},\boldsymbol{\lambda}) \geq 0$.
Then the
condition \eqref{compar1a} is equivalent to the question:
if $\boldsymbol{\sigma}$ is given, then for what
$\boldsymbol{\lambda}$ the following condition holds
\begin{equation}\label{asymp1}
g_{\boldsymbol{\sigma}}(u_{0}) -
g_{\boldsymbol{\sigma}}(v_{0},\boldsymbol{\lambda}) =
o\left(g_{\boldsymbol{\sigma}}(u_{0})\right) \ \mbox{ as } \
g_{\boldsymbol{\sigma}}(u_{0}) \to \infty \ \ ?
\end{equation}

If the condition \eqref{asymp1} is fulfilled and we replace
$\boldsymbol{\sigma}$ by $\boldsymbol{\lambda}$, using the
decision \eqref{defcalA1}--\eqref{defcalA} (oriented on
$\boldsymbol{\sigma}$), then the 1-st kind error probability
$\alpha$ does not change and the 2-nd kind error
probability $\beta_{\boldsymbol{\sigma}}(A,\boldsymbol{\lambda})$
changes slightly.

Generally, the condition \eqref{asymp1} is rather complicated for
checking (since we must find the value $v_{0}$ for each
$\boldsymbol{\lambda}$). Sufficient for having \eqref{asymp1} is
a simpler condition
\begin{equation}\label{asymp1a}
g_{\boldsymbol{\sigma}}(u_{0}) - \max\left\{
g_{\boldsymbol{\sigma}}(u_{0},\boldsymbol{\lambda}),
g_{\boldsymbol{\sigma}}(1,\boldsymbol{\lambda})\right\} =
o\left(g_{\boldsymbol{\sigma}}(u_{0})\right), \qquad
g_{\boldsymbol{\sigma}}(u_{0}) \to \infty
\end{equation}
or, in particular,
\begin{equation}\label{approx3b}
g_{\boldsymbol{\sigma}}(u_{0},\boldsymbol{\lambda}) -
g_{\boldsymbol{\sigma}}(u_{0}) =
\sum\limits_{i=1}^{n}\ln\left[1 +
\frac{u_{0}\sigma_{i}^{2}(\lambda_{i}^{2} -\sigma_{i}^{2})}
{(1+\sigma_{i}^{2})(1+u_{0}\sigma_{i}^{2})}\right] =
o\left(g_{\boldsymbol{\sigma}}(u_{0})\right).
\end{equation}

Note that the condition \eqref{asymp1a} (or \eqref{approx3b})
is only sufficient, but not necessary. It may give satisfactory
results, if $\boldsymbol{\lambda}$ is not very different from
$\boldsymbol{\sigma}$. If $\boldsymbol{\lambda}$ is very different
from $\boldsymbol{\sigma}$, then essential loss of accuracy is
possible (see below example 3, where the condition \eqref{asymp1a}
is not fulfilled, but the condition \eqref{asymp1} is satisfied).
We give another similar example (omitting some details).

E x a m p l e\, 2. Choose $\boldsymbol{\sigma}$ and
$\boldsymbol{\lambda}$, such that $u_{0} \neq v_{0}$, and in the
formula \eqref{asymp1} the equality holds, i.e.
$$
g_{\boldsymbol{\sigma}}(u_{0}) = \max\limits_{u \geq 0}
g_{\boldsymbol{\sigma}}(u) =
g_{\boldsymbol{\lambda}}(v_{0})= \max\limits_{v \geq 0}
g_{\boldsymbol{\lambda}}(v).
$$
Now if the condition \eqref{asymp1a} is satisfied, then
similar condition
\begin{equation}\label{asymp11a}
g_{\boldsymbol{\lambda}}(v_{0}) \leq
g_{\boldsymbol{\sigma}}(v_{0})
\end{equation}
can not be fulfilled. It means that when changing mutually
$\boldsymbol{\sigma}$ and $\boldsymbol{\lambda}$ the condition
\eqref{asymp11a} stops being necessary.

{\bf 2. Case $u_{0} \approx 1$}. Consider an important particular
case when the set ${\mathcal E}$ can be replaced by a point
$\boldsymbol{\sigma} \in {\mathcal E}$, and the sufficient
condition \eqref{approx3b} takes a simple form. Let $\alpha$ be
not very small and we need only that $\alpha(A,{\mathcal E})$
satisfies the inequality
$\alpha(A,{\mathcal E}) \leq 6B_{n}^{-1/2}$, where
$B_{n} = \inf\limits_{\boldsymbol{\sigma} \in {\mathcal E}}
B(\boldsymbol{\sigma})$, and $B(\boldsymbol{\sigma})$ is defined in
\eqref{defD}. In order to do so, keeping in mind some
$\boldsymbol{\sigma} \in {\mathcal E}$, we set
(see \eqref{defD} and \eqref{CLT4})
$$
\begin{gathered}
A = T(\boldsymbol{\sigma})-D(\boldsymbol{\sigma}) + \varepsilon,
\end{gathered}
$$
where
$$
\begin{gathered}
\varepsilon =
\sqrt{B(\boldsymbol{\sigma})\ln B(\boldsymbol{\sigma})}.
\end{gathered}
$$
Denoting $u_{0} = 1-\delta$, $\delta \geq 0$, we show that the value
$\delta$ is small for large $B(\boldsymbol{\sigma})$. Indeed,
the equation \eqref{approx3a1} takes the form
$$
\sum\limits_{i=1}^{n}
\frac{\sigma_{i}^{2}}{1+u_{0}\sigma_{i}^{2}} =
D(\boldsymbol{\sigma})+A = T(\boldsymbol{\sigma}) +
\varepsilon = \sum\limits_{i=1}^{n}
\dfrac{\sigma_{i}^{2}}{1+\sigma_{i}^{2}} + \varepsilon,
$$
from which it follows
$$
\begin{gathered}
\sum\limits_{i=1}^{n}\frac{\delta\sigma_{i}^{4}}
{(1+\sigma_{i}^{2}-\delta\sigma_{i}^{2})(1+\sigma_{i}^{2})} =
\varepsilon \geq
\sum\limits_{i=1}^{n}\frac{\delta\sigma_{i}^{4}}
{(1+\sigma_{i}^{2})^{2}} = \frac{\delta B}{2}.
\end{gathered}
$$
Therefore
$$
\begin{gathered}
0 \leq 1-u_{0} = \delta \leq \frac{2\varepsilon}{B} =
2\sqrt{\frac{\ln B}{B}}.
\end{gathered}
$$
Since $g_{\boldsymbol{\sigma}}(1) = -A/2$,
$g'_{\boldsymbol{\sigma}}(1) = -\varepsilon/2$ and
$g_{\boldsymbol{\sigma}}''(u) < 0$, then
$$
\begin{gathered}
-A \leq 2g_{\boldsymbol{\sigma}}(u_{0}) =
2g_{\boldsymbol{\sigma}}(1-\delta) \leq
2g_{\boldsymbol{\sigma}}(1)-2\delta g'_{\boldsymbol{\sigma}}(1)
= -A + \delta\varepsilon \leq -A + 2\ln B.
\end{gathered}
$$
Therefore, in the sufficient condition \eqref{approx3b} we may set
$u_{0} = 1$ and then it takes the form
\begin{equation}\label{approx3c}
g_{\boldsymbol{\sigma}}(1,\boldsymbol{\lambda}) -
g_{\boldsymbol{\sigma}}(1) =
\sum\limits_{i=1}^{n}\ln\left[1 +
\frac{\sigma_{i}^{2}(\lambda_{i}^{2} -\sigma_{i}^{2})}
{(1+\sigma_{i}^{2})^{2}}\right] =
o\left(g_{\boldsymbol{\sigma}}(1)\right),
\qquad g_{\boldsymbol{\sigma}}(1) \to \infty.
\end{equation}

The results obtained can be formulated as follows

P r o p o s i t i o n \,3. 1) {\it If there exists
$\boldsymbol{\sigma} \in {\mathcal E}$ such that for any
$\boldsymbol{\lambda} \in {\mathcal E}$ the condition
\eqref{approx3b} is satisfied then the property \eqref{compar1a}
holds and the set ${\mathcal E}$ can be replaced by the point
$\boldsymbol{\sigma} \in {\mathcal E}$
without any loss of detection quality}.

2) {\it If only
$\alpha(A,{\mathcal E}) \leq 6B_{n}^{-1/2}$ is desirable and there
exists $\boldsymbol{\sigma} \in {\mathcal E}$ such that for any
$\boldsymbol{\lambda} \in {\mathcal E}$ the condition
\eqref{approx3c} is satisfied then the property \eqref{compar1a}
holds and the set ${\mathcal E}$ can be replaced by the point
$\boldsymbol{\sigma} \in {\mathcal E}$
without any loss of detection quality}.

In the case of stationary sequences similar to
\eqref{approx3c} condition appeared from different arguments in
\cite[Theorem 1, formula (6)]{ZhangPoor11}. Authors of
\cite{ZhangPoor11} called the analog of the condition
\eqref{approx3c} ``surprising'' since, in particular, it does not
demand the set ${\mathcal E}$ to be convex. But, as was already
mentioned (see Remark 2), in the considered problems with
unknown correlations such convexity is not so important.
The condition \eqref{approx3c} itself is a corollary of a quite
natural sufficient condition \eqref{asymp1a}.

It is shown in Appendix that under certain assumptions
upper bounds for $\beta(A,\boldsymbol{\sigma})$ and
$\beta_{\boldsymbol{\sigma}}(A,\boldsymbol{\lambda})$ used above
give exact logarithmic asymptotics for them as $n \to \infty$
(and then it is sufficient to compare functions
$g_{\boldsymbol{\sigma}}(u)$ and
$g_{\boldsymbol{\sigma}}(v,\boldsymbol{\lambda})$).

\begin{center}
{\large\bf \S\,5. Relation of $A$ and
$\alpha(A,\boldsymbol{\sigma})$}
\end{center}

{\bf 1. Central Limit Theorem}. If the given value
$\alpha(A,\boldsymbol{\sigma})$ is not too small, then it is
possible to evaluate it rather accurately using the
Central Limit Theorem and Berry--Esseen inequality.
Let $X_{1}, \ldots,X_{n}$ - independent random variables,
$\mathbf{E}X_{j} = 0$, $\mathbf{E}|X_{j}|^{3} < \infty$,
$j=1,\ldots,n$. Denote
$$
\begin{gathered}
b_{j}^{2} = \mathbf{E}X_{j}^{2}, \qquad
B_{n} = \sum\limits_{j=1}^{n}b_{j}^{2}, \qquad
F_{n}(x) = \mathbf{P}\left(B_{n}^{-1/2}
\sum\limits_{j=1}^{n}X_{j} < x\right),   \\
L_{n} = B_{n}^{-3/2}\sum\limits_{j=1}^{n}\mathbf{E}|X_{j}|^{3}.
\end{gathered}
$$
Then by Berry--Esseen inequality \cite[Ch. V, \S 2, Theorem 3]{P}
$$
\begin{gathered}
\sup_{x}\left|F_{n}(x) - \Phi(x)\right| \leq L_{n}.
\end{gathered}
$$
In our case
$$
\begin{gathered}
X_{j} = \frac{\sigma_{j}^{2}(\xi_{j}^{2}-1)}{1+\sigma_{j}^{2}},
\qquad b_{j}^{2} = \frac{2\sigma_{j}^{4}}{(1+\sigma_{j}^{2})^{2}},
\qquad B_{n} = \sum\limits_{j=1}^{n}b_{j}^{2}, \\
\mathbf{E}|X_{j}|^{3} \leq
\frac{10\sigma_{j}^{6}}{(1+\sigma_{j}^{2})^{3}} \leq
\frac{10\sigma_{j}^{4}}{(1+\sigma_{j}^{2})^{2}} = 5b_{j}^{2},
\qquad L_{n} \leq 5B_{n}^{-1/2}.
\end{gathered}
$$
Therefore
$$
\begin{gathered}
\alpha(A,\boldsymbol{\sigma}) =
\mathbf{P}\left(\sum_{i=1}^{n}\frac{\sigma_{i}^{2}\xi_{i}^{2}}
{1+\sigma_{i}^{2}} > D(\boldsymbol{\sigma}) +A\right) =
\mathbf{P}\left(\sum_{i=1}^{n}X_{i} > D(\boldsymbol{\sigma}) +A-
T(\boldsymbol{\sigma})\right)
\end{gathered}
$$
and then we get
$$
\begin{gathered}
\left|\alpha(A,\boldsymbol{\sigma})- Q(x)\right| \leq
5B_{n}^{-1/2}, \qquad
x = B_{n}^{-1/2}\left[D(\boldsymbol{\sigma})+A -
T(\boldsymbol{\sigma})\right],
\end{gathered}
$$
where
$$
Q(x) = \frac{1}{\sqrt{2\pi}}\int\limits_{x}^{\infty}e^{-u^{2}/2}du
\leq \min\left\{\frac{1}{2}, \frac{1}{x\sqrt{2\pi}}\right\}
e^{-x^{2}/2}, \qquad x > 0.
$$
In particular,
\begin{equation}\label{CLT3}
\begin{gathered}
\alpha(A,\boldsymbol{\sigma}) \leq \frac{5}{\sqrt{B}} +
\min\left\{\frac{1}{2}, \frac{1}{z}\sqrt{
\frac{B}{2\pi}}\right\}e^{-z^{2}/(2B)}, \\
z = D(\boldsymbol{\sigma})+A - T(\boldsymbol{\sigma}) > 0,
\end{gathered}
\end{equation}
where $B= B(\boldsymbol{\sigma})$ is defined in \eqref{defD}.
It follows from \eqref{CLT3}

P r o p o s i t i o n \,4. {\it If
$A \geq T(\boldsymbol{\sigma}) -D(\boldsymbol{\sigma})+
\sqrt{B(\ln B -\ln\ln B)}$, then}
\begin{equation}\label{CLT4}
\begin{gathered}
\alpha(A,\boldsymbol{\sigma}) \leq
\frac{6}{\sqrt{B(\boldsymbol{\sigma})}}, \qquad
B = B(\boldsymbol{\sigma}).
\end{gathered}
\end{equation}

The estimate \eqref{CLT4} quite accurately shows dependence of the
value $\alpha(A,\boldsymbol{\sigma})$ on
$B(\boldsymbol{\sigma})$ (for large $B$), if the given value
$\alpha > 5B^{-1/2}$. Usually, $B(\boldsymbol{\sigma}) \sim n$.

{\bf 2. Large deviations. Upper bound}. Since
\begin{equation}\label{defalpha1}
\alpha(A,\boldsymbol{\sigma}) =
\mathbf{P}\left(\sum_{i=1}^{n}r_{i}^{2}\xi_{i}^{2} >
D(\boldsymbol{\sigma})+A\right),
\qquad r_{i}^{2} = \frac{\sigma_{i}^{2}}{1+\sigma_{i}^{2}},
\end{equation}
then for any $t \geq 0$ similarly to \eqref{ineq1a}, \eqref{ineq1b}
we have
$$
\begin{gathered}
\alpha(A,\boldsymbol{\sigma}) \leq
e^{-t[D(\boldsymbol{\sigma})+A]/2}\mathbf{E}
e^{t\sum\limits_{i=1}^{n}r_{i}^{2}\xi_{i}^{2}/2} =
e^{-f_{\boldsymbol{\sigma}}(t)},
\end{gathered}
$$
where
\begin{equation}\label{ineq1bb}
\begin{gathered}
2f_{\boldsymbol{\sigma}}(t) = t[D(\boldsymbol{\sigma})+A] +
\sum\limits_{i=1}^{n}\ln(1-t r_{i}^{2}), \quad
2f_{\boldsymbol{\sigma}}'(t) =
D(\boldsymbol{\sigma})+A -\sum\limits_{i=1}^{n}
\frac{r_{i}^{2}}{1-t r_{i}^{2}}, \\
f_{\boldsymbol{\sigma}}''(t) < 0.
\end{gathered}
\end{equation}
Since both conditions \eqref{defA} supposed to be satisfied then
$f_{\boldsymbol{\sigma}}'(0) > 0$ and
$f_{\boldsymbol{\sigma}}'(1) < 0$. Therefore
$\max\limits_{t \geq 0}f_{\boldsymbol{\sigma}}(t)$ is attained for
$0 < t_{0} < 1$, which is determined by the equation
$$
\sum\limits_{i=1}^{n}\frac{r_{i}^{2}}{1-t_{0}r_{i}^{2}} =
\sum\limits_{i=1}^{n}
\frac{\sigma_{i}^{2}}{1+(1-t_{0})\sigma_{i}^{2}}=
D(\boldsymbol{\sigma})+A.
$$
Then
\begin{equation}\label{approx3a33}
\alpha(A,\boldsymbol{\sigma}) \leq
e^{-f_{\boldsymbol{\sigma}}(t_{0})}.
\end{equation}

For $t=1$ we have $f_{\boldsymbol{\sigma}}(1) = A/2$, from which
the estimate follows
\begin{equation}\label{approx3a3b}
\alpha(A,\boldsymbol{\sigma}) \leq
e^{-f_{\boldsymbol{\sigma}}(1)} =e^{-A/2}.
\end{equation}
Simple estimate \eqref{approx3a3b} is sufficiently accurate, if
$t_{0}$ is close to $1$ (i.e. if all $\{\sigma_{i}^{2}\}$ are small).


\begin{center} {\large\bf \S\,6. One more example} \end{center}

Consider more complicated

E x a m p l e\, 3. Let for a given $R > 0$
$$
{\mathcal E} = \left\{\boldsymbol{\sigma} \geq {\mathbf 0}:
\sum_{i=1}^{n}\sigma_{i}^{2} \geq nR^{2}\right\}.
$$
Then
$$
{\mathcal E}_{0} = \left\{\boldsymbol{\sigma} \geq {\mathbf 0}:
\sum_{i=1}^{n}\sigma_{i}^{2} = nR^{2}\right\}.
$$
Denote $\boldsymbol{\sigma}_{0} = (R,\ldots,R)$. Then
$$
\begin{gathered}
D(\boldsymbol{\sigma}_{0}) = n\ln(1+R^{2}), \qquad
\min\limits_{\boldsymbol{\sigma} \in {\mathcal E}}
D(\boldsymbol{\sigma}) = \ln(1+nR^{2})
\end{gathered}
$$
and that minimum is attained for $\boldsymbol{\sigma}$, which has
only one nonzero (equal to $R\sqrt{n}$) coordinate. Denote
$\boldsymbol{\sigma}_{i}$, $i=1,\ldots,n$ all those vectors.
For example, $\boldsymbol{\sigma}_{1} =(R\sqrt{n},0,\ldots,0)$.
Denote also
$$
{\mathcal E}_{1} = \left\{\boldsymbol{\sigma}_{i}, \ i=1,\ldots,n
\right\}.
$$
We show that without any loss of quality (in the sense of
asymptotic equality \eqref{compar1a}) all set ${\mathcal E}$
can be replaced by the set ${\mathcal E}_{1}$ and get the same
results as for one point $\boldsymbol{\sigma}_{1}$. Notice that
it does not follow from the sufficient condition \eqref{asymp1a}.

In order to show possibility of such reduction of the set
${\mathcal E}$ we use the likelihood ratio criteria with the set
${\mathcal E}_{1}$ (see \eqref{defmaxrat1a})
$$
{\mathcal A}(A,{\mathcal E},{\mathcal E}_{1}) = \left\{
{\mathbf y}: 2\sup_{\boldsymbol{\lambda} \in {\mathcal E}_{1}}
r({\mathbf y},\boldsymbol{\lambda}) \leq A \right\}.
$$

Now, if $D(\boldsymbol{\sigma}_{1})+ A=\ln(1+nR^{2})+ A \geq 0$,
then
\begin{equation}\label{defbetaup18}
\begin{gathered}
\beta(A,\boldsymbol{\sigma}_{1}) = \mathbf{P}
\left\{2\max_{\boldsymbol{\lambda} \in {\mathcal E}_{1}}
r({\mathbf y},\boldsymbol{\lambda}) \leq A|\boldsymbol{\sigma}_{1}
\right\} = \\
= \mathbf{P}\left\{2r({\mathbf y},\boldsymbol{\sigma}_{1}) \leq A|
\boldsymbol{\sigma}_{1}\right\}\prod_{i=2}^{n}
\mathbf{P}\left\{2r({\mathbf y},\boldsymbol{\sigma}_{i}) \leq A|
\boldsymbol{\sigma}_{1}\right\} = \\
= \mathbf{P}\left\{\xi_{1}^{2} \leq
\frac{D(\boldsymbol{\sigma}_{1}) + A}{nR^{2}}\right\}
\prod_{i=2}^{n}\mathbf{P}\left\{\frac{nR^{2}\xi_{i}^{2}}
{1+nR^{2}} \leq D(\boldsymbol{\sigma}_{1}) + A\right\} \leq
\frac{\sqrt{D(\boldsymbol{\sigma}_{1}) + A}}{R\sqrt{n}}.
\end{gathered}
\end{equation}
The estimate \eqref{defbetaup18} gives the correct asymptotics in
$n$, since for $n \to \infty$ and small $\alpha(A)$
$$
\begin{gathered}
\prod_{i=2}^{n}\mathbf{P}\left\{\frac{nR^{2}\xi_{i}^{2}}
{1+nR^{2}} \leq D(\boldsymbol{\sigma}_{1}) + A\right\} \sim
\mathbf{P}^{n}\left\{|\xi_{1}| \leq
\sqrt{D(\boldsymbol{\sigma}_{1}) + A}\right\} \approx \\
\approx \left[1-\frac{\alpha(A)}{n}\right]^{n} \approx
e^{-\alpha(A)}.
\end{gathered}
$$

We also have (see estimates \eqref{largegaus1})
$$
\begin{gathered}
\alpha(A) = \mathbf{P}
\left\{2\max_{\boldsymbol{\lambda} \in {\mathcal E}_{1}}
r({\mathbf y},\boldsymbol{\lambda}) \geq A|\mathcal H_{0}\right\}
\leq \mathbf{P}\left\{\max_{i =1,\ldots,n}\xi_{i}^{2}
\geq D(\boldsymbol{\sigma}_{1})+ A\right\} \leq \\
\leq n\mathbf{P}\left\{\xi_{1}^{2} \geq
D(\boldsymbol{\sigma}_{1})+ A\right\} \leq
\frac{n}{\sqrt{D(\boldsymbol{\sigma}_{1})+ A}}
\exp\left\{-\frac{[D(\boldsymbol{\sigma}_{1})+ A]}{2}\right\}.
\end{gathered}
$$
To simplify formulas we set $A$ as follows
$$
\begin{gathered}
A = 2\ln n - D(\boldsymbol{\sigma}_{1}) = 2\ln n -\ln(1+nR^{2}).
\end{gathered}
$$
Then
\begin{equation}\label{defbetaup22a}
\alpha(\boldsymbol{\sigma}_{1}) \leq \frac{1}{\sqrt{2\ln n}}
\qquad \mbox{and} \qquad \beta(\boldsymbol{\sigma}_{1}) \leq
\frac{\sqrt{2\ln n}}{R\sqrt{n}}.
\end{equation}

Consider the value $\beta(\boldsymbol{\lambda})$ for
$\boldsymbol{\lambda} \in {\mathcal E}_{0}$. Denote
$$
z^{2} = \frac{2(1+nR^{2})\ln n}{nR^{2}}.
$$
Without loss of generality, assume
$\lambda_{1} \leq \lambda_{2} \leq \ldots \leq \lambda_{n}$, and
introduce an auxiliary level $\lambda_{0}$,
$0 \leq \lambda_{0} \leq z$
(level $\lambda_{0}$ will be defined below). Then we have
\begin{equation}\label{defbetaup1e30}
\begin{gathered}
\ln \beta(A,{\mathcal E}_{1},\boldsymbol{\lambda}) =
\ln\mathbf{P}\left\{\max_{i=1,\ldots,n}
\left[(1+\lambda_{i}^{2})\xi_{i}^{2}\right] \leq
z^{2}\right\} = B_{1} + B_{2} + B_{3},
\end{gathered}
\end{equation}
where
\begin{equation}\label{defbetaup1e3}
\begin{gathered}
B_{1} = \ln\mathbf{P}\left\{\max_{\lambda_{i} \leq \lambda_{0}}
\left[(1+\lambda_{i}^{2})\xi_{i}^{2}\right] \leq z^{2}\right\}< 0,
\\ B_{2} = \ln\mathbf{P}\left\{
\max_{\lambda_{0}^{2} < \lambda_{i}^{2} \leq z^{2}-1}
\left[(1+\lambda_{i}^{2})\xi_{i}^{2}\right] \leq z^{2}\right\}<0, \\
B_{3} = \ln\mathbf{P}\left\{\max_{\lambda_{i}^{2} > z^{2}-1}
\left[(1+\lambda_{i}^{2})\xi_{i}^{2}\right] \leq z^{2}\right\}<0.
\end{gathered}
\end{equation}
We estimate sequentially values $B_{1},B_{2},B_{3}$ from
\eqref{defbetaup1e3}. For that purpose denote
\begin{equation}\label{defn1n2}
\begin{gathered}
n_{1} = \#\{\lambda_{i}: \lambda_{i} \leq \lambda_{0}\}, \quad
n_{2} = \#\{\lambda_{i}:
\lambda_{0}^{2} < \lambda_{i}^{2} \leq z^{2}-1\}, \\
n_{3} = \#\{\lambda_{i}: \lambda_{i}^{2} > z^{2}-1\}, \\
s_{1}n_{1} = \sum_{\{\lambda_{i} \leq \lambda_{0}\}}\lambda_{i}^{2},
\qquad s_{2}n_{2} =
\sum_{\{\lambda_{0}^{2} < \lambda_{i}^{2} \leq z^{2}-1\}}
\lambda_{i}^{2}, \qquad  s_{3}n_{3} =
\sum_{\{\lambda_{i}^{2} > z^{2}-1\}}\lambda_{i}^{2}.
\end{gathered}
\end{equation}

Using notations \eqref{defn1n2} and the inequality
$\ln(1+z) \leq z$, for the value $B_{1}$ we have
\begin{equation}\label{defbetaup1e31}
\begin{gathered}
B_{1} = \sum_{i=1}^{n_{1}}\ln\left[1 -\mathbf{P}\left\{
(1+\lambda_{i}^{2})\xi_{i}^{2} \geq z^{2}\right\}\right]  \leq \\
\leq -\sum_{i=1}^{n_{1}}\mathbf{P}\left\{
\left(1+\lambda_{i}^{2}\right)\xi_{i}^{2} \geq z^{2}\right\} =
-2\sum_{i=1}^{n_{1}}\mathbf{P}
\left(\sqrt{1+\lambda_{i}^{2}}\,\xi \geq z\right).
\end{gathered}
\end{equation}
Consider first the variational task of minimization of the sum
of any two terms from the right-hand side of \eqref{defbetaup1e31}
$$
\mathbf{P}\left(\sqrt{1+\lambda_{i}^{2}}\,\xi \geq z\right) +
\mathbf{P}\left(\sqrt{1+\lambda_{j}^{2}}\,\xi \geq z\right)
$$
over variables $\lambda_{i},\lambda_{j}$ with a given sum
$\lambda_{i}^{2}+\lambda_{j}^{2} = r^{2}-2 \geq 0$. Denoting by
$u^{2} = 1+\lambda_{i}^{2}$, $v^{2} = 1+\lambda_{j}^{2}$,
$u^{2}+v^{2} = r^{2}$, and by $f(u,z,r)$ that sum, we have
$$
\begin{gathered}
f(u,z,r) = \mathbf{P}\left\{\xi \geq \frac{z}{u}\right\} +
\mathbf{P}\left\{\xi \geq \frac{z}{v}\right\} =
2\left[1-\Phi\left(\frac{z}{u}\right) -
\Phi\left(\frac{z}{v}\right)\right], \quad
v=\sqrt{r^{2}-u^{2}} \geq 1.
\end{gathered}
$$
We are interested in
$\min\limits_{1 \leq u \leq \sqrt{r^{2}-1}}f(u,z,r)$. We have
$$
\begin{gathered}
f'(u) = f'_{u} + f'_{v}v'_{u}, \quad
f'_{u} = \frac{2ze^{-z^{2}/(2u^{2})}}{\sqrt{2\pi}u^{2}}, \quad
f'_{v} = \frac{2ze^{-z^{2}/(2v^{2})}}{\sqrt{2\pi}v^{2}},
\quad v'_{u} = -\frac{u}{v}, \\
f'(u) = \frac{2z}{\sqrt{2\pi}}\left\{
\frac{e^{-z^{2}/(2u^{2})}}{u^{2}}- \frac{ue^{-z^{2}/(2v^{2})}}
{v^{3}}\right\} = \frac{2u}{\sqrt{2\pi}z^{2}}\left[
\exp\left\{t\left(\frac{z}{u}\right)\right\} -
\exp\left\{t\left(\frac{z}{v}\right)\right\}\right],
\end{gathered}
$$
where we denoted
$$
t(x) = 3\ln x - \frac{x^{2}}{2}, \qquad
t'(x) = \frac{3}{x} - x, \qquad x > 0.
$$
The function $t(x)$ monotonically decreases for  $x > \sqrt{3}$
and monotonically increases for $0 < x < \sqrt{3}$. Without loss of
generality we may assume that $z/v \leq z/u$, i.e.
$2u^{2} \leq r^{2}$. Therefore if $z/v = z/\sqrt{r^{2}-u^{2}} \geq
z/\sqrt{r^{2}-1} \geq \sqrt{3}$, then $f'(u) \leq 0$,
$u \leq r/\sqrt{2}$, and then the minimum of the function
$f(u,z,r)$ in $u$ is attained for $u=v$ (i.e. for
$\lambda_{i}=\lambda_{j}$). To fulfill those conditions
it is sufficient to set $r^{2} = z^{2}/3+1$. As a result, we get
that if among $\{\lambda_{i}\}$ there is a pair
$\lambda_{i},\lambda_{j}$ such that $\lambda_{i} \neq \lambda_{j}$
and $\lambda_{i}^{2}+\lambda_{j}^{2} \leq r^{2}-2 = z^{2}/3-1$,
then $f(u,z,r)$ decreases if we replace each
$\lambda_{i}^{2}, \lambda_{j}^{2}$ by their half-sum
$(\lambda_{i}^{2}+ \lambda_{j}^{2})/2$. Therefore we define
the level $\lambda_{0}$ as follows
\begin{equation}\label{defsigma0}
\begin{gathered}
\lambda_{0}^{2} = z^{2}/6-1/2.
\end{gathered}
\end{equation}
Continuing that process of maximization of the right-hand side
of \eqref{defbetaup1e31} we get that its maximum is attained when
$$
\lambda_{1}^{2} =\ldots = \lambda_{n_{1}}^{2} \leq \lambda_{0}^{2}.
$$
For remaining $n-n_{1}$ components $\{\lambda_{i}\}$ we have
$$
\lambda_{i} > \lambda_{0}, \qquad i=n_{1}+1,\ldots,n.
$$
Therefore for the value $B_{1}$ from \eqref{defbetaup1e31}
we get for large $n$ and some $C > 0$
(see estimates \eqref{largegaus1})
\begin{equation}\label{defbetaup1e32}
\begin{gathered}
B_{1} \leq -2n_{1}\mathbf{P}\left(\sqrt{1+\lambda_{1}^{2}}\,
\xi \geq z\right)
\leq -\frac{Cn_{1}}{z}e^{-\frac{z^{2}}{2(1+\lambda_{1}^{2})}} \leq
-\frac{Cn^{\frac{\lambda_{1}^{2}}{1+\lambda_{1}^{2}}}}
{\sqrt{\ln n}},
\end{gathered}
\end{equation}
since $(n-n_{1})\lambda_{0}^{2} \leq R^{2}n$ and then
$n_{1} \geq n\left(1-R^{2}/\lambda_{0}^{2}\right).$

For the value $B_{2}$ from \eqref{defbetaup1e3} we get
\begin{equation}\label{defbetaup1e2bb}
\begin{gathered}
B_{2} \leq n_{2}\ln\mathbf{P}\left\{(1+\lambda_{0}^{2})\xi^{2}
\leq z^{2}\right\} \leq n_{2}\ln\mathbf{P}
\left\{|\xi| \leq \sqrt{6}\right\} \leq - \frac{n_{2}}{100}.
\end{gathered}
\end{equation}
Now we estimate the value $B_{3}$ from \eqref{defbetaup1e3}.
We have
\begin{equation}\label{defbetaup1e2b}
\begin{gathered}
B_{3} = \sum_{\lambda_{i}^{2} > z^{2}-1}\ln\left(2\mathbf{P}\left\{
0 \leq \xi_{i} \leq \frac{z}{\sqrt{1+\lambda_{i}^{2}}}\right\}
\right) \leq -\frac{n_{3}}{2}\ln\frac{\pi}{2} - \frac{1}{2}I_{3},
\end{gathered}
\end{equation}
where
$$
I_{3} = \sum_{\{\lambda_{i}^{2} \geq z^{2}-1\}}
\ln\frac{1+\lambda_{i}^{2}}{z^{2}}.
$$
Consider the value $I_{3}$ for given $s_{3}$ and $n_{3}$.
Since the function $I_{3}$ is $\bigcap$-concave in
$\{\lambda_{i}^{2}\}$, its minimum is attained at an extreme point,
i.e. when one of coordinates $\lambda_{j}^{2}$ equals
$s_{3}n_{3}-(n_{3}-1)(z^{2}-1)$, and all remaining coordinates
$\lambda_{i}^{2}$ equal $z^{2}-1$. Hence
\begin{equation}\label{defbetaup26a}
\begin{gathered}
I_{3} \geq \ln\frac{1+s_{3}n_{3}-(n_{3}-1)(z^{2}-1)}{z^{2}} =
\ln\frac{z^{2}+n_{3}(s_{3}-z^{2}+1)}{z^{2}}.
\end{gathered}
\end{equation}

Therefore from \eqref{defbetaup1e30} and
\eqref{defbetaup1e32}--\eqref{defbetaup26a} we get for large $n$
\begin{equation}\label{defbetaup1e302}
\begin{gathered}
\ln \beta(A,{\mathcal E}_{1},\boldsymbol{\lambda})
\leq -\frac{Cn^{\frac{\lambda_{1}^{2}}{1+\lambda_{1}^{2}}}}
{\sqrt{\ln n}} -\frac{n_{2}}{100} -\frac{n_{3}}{5} - \frac{1}{2}
\ln\frac{z^{2}+n_{3}(s_{3}-z^{2}+1)}{z^{2}}.
\end{gathered}
\end{equation}

It remains to show that the right-hand side of
\eqref{defbetaup1e302} satisfies the inequality
\begin{equation}\label{defbetaup1e303}
\begin{gathered}
\min_{n_{2},n_{3},\lambda_{1}}\left\{
\frac{Cn^{\frac{\lambda_{1}^{2}}{1+\lambda_{1}^{2}}}}
{\sqrt{\ln n}} + \frac{n_{2}}{100} +\frac{n_{3}}{5} +
\frac{1}{2}\ln\frac{z^{2}+n_{3}(s_{3}-z^{2}+1)}{z^{2}}\right\} \geq
\\ \geq \frac{1}{2}\ln(R^{2}n) + o(\ln(R^{2}n)),
\end{gathered}
\end{equation}
where minimum is taken provided
$n_{2}\lambda_{0}^{2} + n_{3}s_{3} \geq R^{2}n + o(R^{2}n)$.

We may assume that
(see \eqref{defbetaup22a} and \eqref{defbetaup1e32})
$$
\begin{gathered}
n_{2} < 50\ln(R^{2}n), \qquad
n_{3} < 3\ln(R^{2}n) \qquad \mbox{ и }
\qquad \lambda_{1}^{2} < \frac{2\ln\ln n}{\ln n} +
\frac{2\ln R}{\ln^{2}n}
\end{gathered}
$$
(otherwise the inequality \eqref{defbetaup1e303} holds). In other
words, almost all power $R^{2}n$ is distributed on last $n_{3}$
components. Hence
$$
\begin{gathered}
n_{3}s_{3} = R^{2}n -n_{1}\lambda_{1}^{2}- n_{2}z^{2} =
R^{2}n + o(n).
\end{gathered}
$$
Then the inequality \eqref{defbetaup1e303} holds and therefore for
any $\boldsymbol{\lambda} \in {\mathcal E}$ we get as
$n \to \infty$
\begin{equation}\label{defbetaup1e33}
\begin{gathered}
\ln \beta(A,{\mathcal E}_{1},\boldsymbol{\lambda}) \leq
-\frac{1}{2}\ln(R^{2}n) + o(\ln(R^{2}n))
= (1+o(1))\ln\beta(\boldsymbol{\sigma}_{1}).
\end{gathered}
\end{equation}
The relation \eqref{defbetaup1e33} means that the likelihood
ratio criteria with the set ${\mathcal E}_{1}$ allows to get
for the whole set ${\mathcal E}$ the same results as for the single
point $\boldsymbol{\sigma}_{1}$.


\hfill {\large\sl APPENDIX}

{\bf 1. Tails of ${\cal N}(0,1)$}. Let
$\xi \sim {\cal N}(0,1)$. Then the following estimates are known
\begin{equation}\label{largegaus1}
\begin{gathered}
\frac{ze^{-z^{2}/2}}{(z^{2}+1)\sqrt{2\pi}} \leq
\mathbf{P}\left\{\xi \geq z\right\} \leq \frac{e^{-z^{2}/2}}
{z\sqrt{2\pi}}, \qquad z > 0,
\end{gathered}
\end{equation}
where the lower bound is derived via integration by parts.

{\bf 2. Distribution $\chi^{2}$. Large deviations}.
Consider the value
\begin{equation}\label{defbeta1}
\beta(A,n) =
\mathbf{P}\left(\sum\limits_{i=1}^{n}\xi_{i}^{2}< A\right).
\end{equation}

L e m m a\, 3. {\it For $A \leq n$ and $n \geq 1$
the following estimates hold}
\begin{equation}\label{lnbeta11}
-\frac{1}{2}\ln(\pi n) - \frac{1}{3n} \leq \ln \beta(A,n) +
\frac{1}{2}\left(n\ln\frac{n}{eA} + A\right) \leq 0.
\end{equation}

{\it Proof}. The right one of inequalities \eqref{lnbeta11}
follows from exponential Chebychev \\ inequality
(see \eqref{ineq18} и \eqref{ineq1a}). To prove
the left one of inequalities \eqref{lnbeta11} denote
\begin{equation}\label{defB1}
\mathcal B_{n}(r) =
\left\{{\mathbf y}: \sum_{i=1}^{n}y_{i}^{2} \leq r^{2}\right\}.
\end{equation}
Then
$$
\begin{gathered}
|\mathcal B_{n}(r)| =\frac{\pi^{n/2}r^{n}}{\Gamma(n/2+1)}, \\
\ln\Gamma(z) = z\ln\frac{z}{e} - \frac{1}{2}\ln z +
\frac{1}{2}\ln(2\pi) +
\frac{\theta}{6z}, \qquad z >  0, \qquad 0 \leq \theta \leq 1.
\end{gathered}
$$
Therefore
$$
\begin{gathered}
\beta(A,n) =\frac{1}{(2\pi)^{n/2}}
\int\limits_{0}^{\sqrt{A}}e^{-r^{2}/2}d|\mathcal B_{n}(r)| =
\frac{1}{\Gamma(n/2)}\int\limits_{0}^{A/2}v^{n/2-1}e^{-v}dv.
\end{gathered}
$$
Integrating by parts, we have ($a=n/2-1$, $B=A/2$, $0 < \theta < 1$)
$$
\begin{gathered}
\int\limits_{0}^{B}v^{a}e^{-v}dv = \frac{B^{a+1}e^{-B}}{a+1} +
\frac{1}{(a+1)}\int\limits_{0}^{B}v^{a+1}e^{-v}dv
= \frac{B^{a+1}e^{-B}}{a+1} +
\frac{\theta B}{(a+1)}\int\limits_{0}^{B}v^{a}e^{-v}dv,
\end{gathered}
$$
Therefore
$$
\begin{gathered}
\int\limits_{0}^{B}v^{a}e^{-v}dv =
\frac{B^{a+1}e^{-B}}{a+1-\theta B}, \qquad
0 < \theta < 1.
\end{gathered}
$$
Then
$$
\begin{gathered}
\beta(A,n) = \frac{1}{\Gamma(n/2)}\int\limits_{0}^{A/2}
v^{n/2-1}e^{-v}dv = \frac{2}{(n-\theta A)\Gamma(n/2)}
\left(\frac{A}{2}\right)^{n/2}e^{-A/2}.
\end{gathered}
$$
Hence ($0 \leq \theta,\theta_{1} \leq 1$)
$$
\begin{gathered}
\ln \beta(A,n) =
-\frac{n}{2}\ln\frac{n}{eA} - \frac{A}{2} +
\frac{1}{2}\ln\frac{n}{4\pi} - \frac{\theta_{1}}{3n}
+ \ln\frac{2}{n-\theta A},
\end{gathered}
$$
from which the left one of inequalities \eqref{lnbeta11} follows.
$\qquad \triangle$

Consider the value
\begin{equation}\label{defalpha11}
\alpha(A,n) =
\mathbf{P}\left(\sum\limits_{i=1}^{n}\xi_{i}^{2}> A\right).
\end{equation}

L e m m a\, 4.
{\it For $A \geq n$ and $n \geq 2$ the following estimates hold}
\begin{equation}\label{lnalpha11}
\begin{gathered}
-\frac{1}{3n} - \frac{1}{2}\ln\frac{\pi A^{2}}{n} \leq
\ln \alpha(A,n) +\frac{1}{2}\left(n\ln\frac{n}{eA} + A\right)
\leq 0.
\end{gathered}
\end{equation}

{\it Proof}. The right one of inequalities \eqref{lnalpha11}
follows from exponential Chebychev \\
inequality
(see \eqref{defalpha1}--\eqref{approx3a33}). To prove
the left one of inequalities \eqref{lnalpha11}, we have, using
the notation \eqref{defB1}
$$
\begin{gathered}
\alpha(A,n) =\frac{1}{(2\pi)^{n/2}}
\int\limits_{\sqrt{A}}^{\infty}e^{-r^{2}/2}d|\mathcal B_{n}(r)| =
\frac{1}{\Gamma(n/2)}\int\limits_{A/2}^{\infty}v^{n/2-1}e^{-v}dv.
\end{gathered}
$$
Integrating by parts, we have for the last integral
($a=n/2-1$, $B=A/2$)
$$
\begin{gathered}
\int\limits_{B}^{\infty}v^{a}e^{-v}dv = B^{a}e^{-B} +
\frac{\theta a}{B}\int\limits_{B}^{\infty}v^{a}e^{-v}dv,
\qquad 0 < \theta < 1.
\end{gathered}
$$
Therefore
$$
\begin{gathered}
\int\limits_{B}^{\infty}v^{a}e^{-v}dv =
\frac{B^{a}e^{-B}}{1-\theta a/B}, \qquad
0 < \theta < 1,
\end{gathered}
$$
and then
$$
\begin{gathered}
\alpha(A,n)= \frac{2}{A\Gamma(n/2)}
\left(\frac{A}{2}\right)^{n/2}e^{-A/2}\frac{1}
{1-\theta(n-2)/A}, \qquad 0 < \theta < 1.
\end{gathered}
$$
As a result, for $A \geq n$ and $n \geq 2$ we get
($0 < \theta, \theta_{1} < 1$)
$$
\begin{gathered}
\ln \alpha(A,n) = -\frac{n}{2}\ln\frac{n}{eA} -
\frac{A}{2} -\frac{1}{2}\ln\frac{\pi}{n} -
\frac{\theta_{1}}{3n} - \ln[A-\theta(n-2)],
\end{gathered}
$$
from which the left one of inequalities \eqref{lnalpha11} follows.
$\qquad \triangle$

{\bf 3. Large deviations for $\beta(A,\boldsymbol{\sigma})$.
Lower bound}. To estimate
$\beta(A,\boldsymbol{\sigma})$ from below, we use the approach,
similar to \cite[proof of Theorem 1]{Chernov1}. Let
$\sigma_{1} \geq \sigma_{2} \geq \ldots \sigma_{n}$. We partition
the segment $[1,n]$ onto $K$ equal parts of length
$\Delta = (n-1)/K$ by points $n_{k} = 1+\Delta k$,
$1 \leq k \leq K$, and represent $A$ as a sum
$A = A_{1} + \ldots + A_{K}$. Then
\begin{equation}\label{low1}
\begin{gathered}
\beta(A,\boldsymbol{\sigma})= \mathbf{P}\left(\sum\limits_{i=1}^{n}
\sigma_{i}^{2}\xi_{i}^{2}< A\right)
\geq
\max \prod_{k=1}^{K}
\mathbf{P}\left(\sigma_{n_{k-1}+1}^{2}
\sum\limits_{i=n_{k-1}+1}^{n_{k}}\xi_{i}^{2}< A_{k}\right),
\end{gathered}
\end{equation}
where maximum is taken over all $K$ and $\{A_{k}\}$. To
evaluate probabilities in the right-hand side of \eqref{low1},
we use the estimate \eqref{lnbeta11}. Denoting
$$
b_{k} = \sigma_{n_{k-1}+1}^{2}, \qquad k=1,\ldots,K
$$
and assuming $A_{k} \leq b_{k}\Delta$, $k=1,\ldots,K$
(see \eqref{defbeta1}), we have from \eqref{lnbeta11}
\begin{equation}\label{low2}
\begin{gathered}
2\ln\beta(A,\boldsymbol{\sigma}) \geq 2\max \sum_{k=1}^{K}
\ln\mathbf{P}\left(\sum\limits_{i=n_{k-1}+1}^{n_{k}}\xi_{i}^{2} <
\frac{A_{k}}{b_{k}}\right) \geq \\
\geq -\min_{\{A_{k}\}}
\sum_{k=1}^{K}\left(\Delta\ln\frac{b_{k}}{A_{k}} +
\frac{A_{k}}{b_{k}}\right) - (n-1)\ln\frac{\Delta}{e} -
K\ln(\pi \Delta).
\end{gathered}
\end{equation}
Minimum in the right-hand side of \eqref{low2} provided
$A = A_{1} + \ldots + A_{K}$ is attained for
$$
A_{k} = \frac{\Delta b_{k}}{1 + u_{1}b_{k}}, \qquad
k=1,\ldots,K,
$$
where $u_{1}$ is determined by the equation similar to
\eqref{approx3a1}
\begin{equation}\label{approx3d1}
\sum_{k=1}^{K}\frac{\Delta b_{k}}{1 + u_{1}b_{k}} = A.
\end{equation}
Since $\Delta b_{k} \geq A_{k}$ then $u_{1} \geq 0$. Moreover,
$$
\begin{gathered}
\sum_{k=1}^{K}\left(\Delta\ln\frac{b_{k}}{A_{k}} +
\frac{A_{k}}{b_{k}}\right) =
\Delta\sum_{k=1}^{K}\ln(1 +u_{1}b_{k}) -
(n-1)\ln \frac{\Delta}{e} -u_{1}A.
\end{gathered}
$$
Since for any $u \geq 0$
$$
\begin{gathered}
\Delta\sum_{k=1}^{K}\ln(1 +u b_{k+1}) \leq
\sum\limits_{i=1}^{n}\ln(1+ u \sigma_{i}^{2}) \leq
\Delta\sum_{k=1}^{K}\ln(1 + u b_{k}),
\end{gathered}
$$
then using \eqref{ineq1b}, we have
\begin{equation}\label{approx3cc}
\begin{gathered}
2\ln \beta(A,\boldsymbol{\sigma}) + K\ln\frac{\pi n}{K} \geq
-\Delta\sum_{k=1}^{K}\ln(1 + u_{1}b_{k}) + u_{1}A = \\
= -2g_{\boldsymbol{\sigma}}(u_{0}) -
\Delta\sum_{k=1}^{K}\ln(1 + u_{1}b_{k}) +
\sum\limits_{i=1}^{n}\ln(1+ u_{0}\sigma_{i}^{2}) +
(u_{1} - u_{0})A \geq \\
\geq -2g_{\boldsymbol{\sigma}}(u_{0}) - \Delta\sum_{k=1}^{K}
\ln\frac{1 + u_{1}b_{k}}{1 + u_{1}b_{k+1}}+ (u_{1}- u_{0})A \geq \\
\geq -2g_{\boldsymbol{\sigma}}(u_{0}) -
\Delta\ln\frac{1 + u_{1}b_{1}}{1 + u_{1}b_{n}} =
-2g_{\boldsymbol{\sigma}}(u_{0}) -\Delta\ln\frac{b_{1}}{b_{n}},
\end{gathered}
\end{equation}
where the inequality $u_{1} \geq u_{0}$ was used. Indeed, from
the formula \eqref{approx3d1} we have
$$
(u_{1})'_{b_{k}} = \frac{1}{b_{k}^{2}} \geq 0, \qquad
k=1,\ldots,K,
$$
and since $\sigma_{1} \geq \ldots \geq \sigma_{n}$, we get that
$u_{1} \geq u_{0}$. Therefore denoting
$$
\delta_{\boldsymbol{\sigma}} =
\ln\frac{\max\limits_{i}\sigma_{i}^{2}}
{\min\limits_{i}\sigma_{i}^{2}} \geq 0,
$$
from \eqref{approx3cc} we get
$$
\begin{gathered}
\ln \beta(A,\boldsymbol{\sigma}) +
g_{\boldsymbol{\sigma}}(u_{0}) \geq -\frac{1}{2}\min_{K \geq 1}
\left\{\frac{n\delta_{\boldsymbol{\sigma}}}{K}
+ K\ln(\pi n)\right\} =
- \left[n\delta_{\boldsymbol{\sigma}}\ln(\pi n)
\right]^{1/2},
\end{gathered}
$$
provided that maximizing $K = K_{0} \geq 1$, where
$$
K_{0}^{2} = \frac{n\delta_{\boldsymbol{\sigma}}}{\ln(\pi n)}.
$$
If $K_{0} < 1$ (i.e. $\sigma_{1}^{2}/\sigma_{n}^{2}$ is close to
$1$), then setting $K=1$ we get
$$
\begin{gathered}
\ln \beta(A,\boldsymbol{\sigma}) +
g_{\boldsymbol{\sigma}}(u_{0}) \geq -\ln(\pi n).
\end{gathered}
$$

Both cases $K_{0} \geq 1$, $K_{0} < 1$, and the formula
\eqref{approx3a3} can be combined as follows
\begin{equation}\label{approx3ce}
\begin{gathered}
-\sqrt{\delta_{\boldsymbol{\sigma}}n\ln(\pi n)} - \ln(\pi n) \leq
\ln \beta(A,\boldsymbol{\sigma}) +
g_{\boldsymbol{\sigma}}(u_{0}) \leq 0.
\end{gathered}
\end{equation}

Notice that usually $g_{\boldsymbol{\sigma}}(u_{0}) \sim n$.
Then \eqref{approx3ce} gives the right logarithmic asymptotics for
$\beta(A,\boldsymbol{\sigma})$, if
$\delta_{\boldsymbol{\sigma}} = o(n/\ln n)$, $n \to \infty$.

As a result, we get

P r o p o s i t i o n \,5. 1) {\it For the value
$\ln\beta(A,\boldsymbol{\sigma})$ upper and lower bounds
\eqref{approx3ce} hold}.

2) {\it If $g_{\boldsymbol{\sigma}}(u_{0}) \leq
g_{\boldsymbol{\lambda}}(u_{0})$}
({\it for example, the sufficient condition \eqref{approx3b}}
is fulfilled),
{\it then}
\begin{equation}\label{theor1}
\begin{gathered}
\ln\beta(A,\boldsymbol{\lambda}) \leq
\ln\beta(A,\boldsymbol{\sigma}) +
\sqrt{\delta_{\boldsymbol{\sigma}} n\ln(\pi n)} +
\ln(\pi n).
\end{gathered}
\end{equation}

The formula \eqref{theor1} follows from \eqref{approx3ce}:
$$
\begin{gathered}
\ln\beta(A,\boldsymbol{\lambda}) \leq
-g_{\boldsymbol{\lambda}}(u_{0}) \leq
-g_{\boldsymbol{\sigma}}(u_{0}) \leq
\ln\beta(A,\boldsymbol{\sigma}) +
\sqrt{\delta_{\boldsymbol{\sigma}} n\ln(\pi n)} + \ln(\pi n).
\end{gathered}
$$

Similarly the lower bound for $\alpha(A,\boldsymbol{\sigma})$
can be derived.


\vspace{5mm}

\begin{flushleft}
{\small {\it Burnashev Marat Valievich} \\
Kharkevich Institute for Information Transmission Problems, \\
Russian Academy of Sciences, Moscow\\
 {\tt burn@iitp.ru}}
\end{flushleft}%


\newpage

\begin{center}{\large  REFERENCES} \end{center}
\begin{enumerate}
\bibitem{Wald}
A. Wald, Statistical Decision Functions, Wiley, New York, 1950.
\bibitem{Leman}
{\it Lehmann E. L.} Testing of Statistical Hypotheses.
New York: Wiley, 1959.
\bibitem{Bur79}
{\it Burnashev M. V.}
On the Minimax Detection of an Inaccurately Known Signal in a
White Gaussian Noise Background //
Theory of Prob. and Its Appl. 1979. V. 24. no. 1. P. 106--118.
\bibitem{ZhangPoor11}
{\it Zhang W., Poor H.V.} On Minimax Robust Detection of Stationary
Gaussian Signals in White Gaussian Noise // IEEE Trans. Inform.
Theory. 2011. V. 57. no. 6. P. 3915--3924.
\bibitem{Ponom85}
{\it Ponomarenko L. S.} On Estimating Distributions of
Normalized Quadratic Forms of Normally Distributed Random
Variables //
Theory of Prob. and Its Appl. 1985. V. 30. no. 3. P. 545--549.
\bibitem{Bak95}
{\it Bakirov N. K.} Comparison Theorems for Distribution
Functions of Quadratic Forms of Gaussian Vectors //
Theory of Prob. and Its Appl. 1995. V. 40. no. 2. P. 404--412.
\bibitem{Bur16}
{\it Burnashev M. V.} Two theorems on distribution of Gaussian
quadratic forms // Problems of Information Transmission. 2017.
Т. 37. № . С. .
\bibitem{Chernov1}
{\it Chernov H.} A Measure of Asymptotic Efficiency for Tests of
a Hypothesis Based on the Sum of Observations // Annals of
Mathematical Statistics. 1952. V. 23. № 6. P. 493--507.
\bibitem{P}
{\it Petrov V. V.} Sums of independent random variables.
Springer, 1975.

\end{enumerate}
\end{document}